\documentclass[%
 aip,
 amsmath,amssymb,
 reprint,
]{revtex4-1}

\usepackage{graphicx}
\usepackage{dcolumn}
\usepackage{bm}

\usepackage[utf8]{inputenc}
\usepackage[T1]{fontenc}
\usepackage{mathptmx}
\usepackage{etoolbox}

\makeatletter
\def\@email#1#2{%
 \endgroup
 \patchcmd{\titleblock@produce}
  {\frontmatter@RRAPformat}
  {\frontmatter@RRAPformat{\produce@RRAP{*#1\href{mailto:#2}{#2}}}\frontmatter@RRAPformat}
  {}{}
}%
\makeatother
\begin{document}

\title[]{Apparatus for Optical-Atomic System Integration \& Calibration: 1 atm to 1$\times$10$^{-11}$ Torr in 24h}

\affiliation{Department of Physics, University of California San Diego, California 92093, USA}
\author{G. Kestler}
\author{K. Ton}
\author{J. T. Barreiro}

\date{\today}

\begin{abstract}
Ultracold atoms exquisitely controlled by lasers are the quantum foundation, particularly for sensing, timekeeping, and computing, of state-of-the-art quantum science and technology. However, the laboratory-scale infrastructure for such optical-atomic quantum apparatuses rarely translates into commercial applications. A promising solution is miniaturizing the optical layouts onto a chip-scale device integrated with cold atoms inside a compact ultra-high vacuum (UHV) chamber. For prototyping purposes, however, rapidly loading or exchanging test photonic devices into a UHV chamber is limited by the evacuation time from atmospheric pressures to the optimal pressures for ultracold atoms of $1\times10^{-11}$~Torr, a process typically taking weeks or months without cryogenics. Here, we present a loadlock apparatus and loading procedure capable of venting, exchanging, and evacuating back to $<1\times10^{-11}$~Torr in under 24 hours. Our system allows for rapid testing and benchmarking of various photonic devices with ultracold atoms.  
\end{abstract}

\maketitle

\section{Introduction}

Optically controlled ultracold atoms are at the core of versatile and powerful platforms for quantum science and technologies. These platforms are used for highly accurate sensors, which often rely on matterwave interferometry~\cite{Parker2018, Moan2020, Panda2024} and precision spectroscopy~\cite{Ye2008, Okaba2014}. Matterwave interferometers have demonstrated substantial improvements for inertial sensing~\cite{Amico2022}, and devices leveraging spectroscopy with narrow-line atomic transitions provide the most stable and accurate atomic clocks to date~\cite{Ludlow2006, Boyd2007, Bloom2014, Nicholson2015, Aeppli2024}. Beyond sensing and timekeeping applications, such experiments also contribute to measurements of fundamental constants~\cite{Parker2018}, collective radiative enhancements through cavity quantum electrodynamics (QED)~\cite{Goban2015, Zhou2024}, and scalable quantum computing architectures with neutral atom qubits~\cite{Saffman2016, Weiss2017, Cooper2018, Madjarov2019}. However, nearly all such experiments rely on laboratory-scale complex optical setups and ultra-high vacuum (UHV) chambers with limited optical access. Miniaturization by integrating cold atoms with an optical setup on a chip~\cite{Hummon2018, Kitching2018} would make quantum technologies more accessible for quantum sensing, atomic timekeeping, and quantum computing.

Developing novel complex technologies is an iterative process involving extensive prototyping, benchmarking, and validation. In particular, the UHV pressures required for cold atom experiments limit rapid testing since each time a new device is loaded, the chamber is subjected to atmospheric pressures. One solution is to use specially designed UHV loadlock systems, where devices are loaded into a separate chamber called a loadlock, which can be rapidly evacuated to UHV pressures before transferring the device to a science chamber with ultracold atoms. So far, this process takes about a week to vent and return to $5\times10^{-10}$~Torr~\cite{Yin2023}. Though using cryogenics to cool the loadlock and science chamber significantly speeds up the process~\cite{Visser1981, Chottiner1987}, this approach requires external cooling apparatuses to maintain UHV pressures.

Here, we present a loadlock-based apparatus and loading procedure capable of opening the chamber to atmospheric pressures and returning to a UHV of $1\times10^{-11}$~Torr within 24 hours. The entire apparatus sits on a $30"\times48"$ optical breadboard, and the $\approx$1.3~L total loadlock volume also allows for isolation of the experiment with only ion pumps and non-evaporable getters (NEGs). 

\section{Vacuum Apparatus}

\begin{figure*}[t!]
    \includegraphics{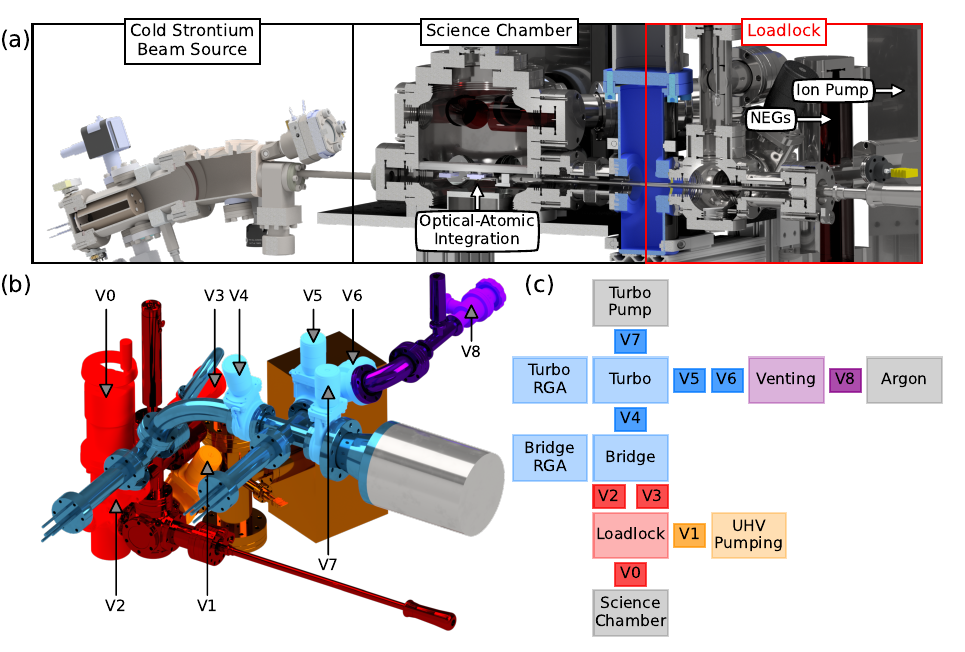}
    \caption{(a) The complete experimental apparatus consists of a cold atom source, a science chamber with ultracold atoms, and a loadlock chamber. An all-metal VAT gate valve separates the loadlock chamber and science chamber. Samples are loaded into the loadlock chamber at atmospheric pressures and evacuated to $<1\times10^{-11}$~Torr before transferring to the science chamber, which is kept at $1\times10^{-11}$~Torr. (b,c) The loadlock chamber is further split into four subsections: {\em loadlock} (red), {\em UHV pumps} (orange), {\em bridge \& turbo} (blue), {\em venting} (purple). Valves V0-V8 are used to ensure a controlled and clean venting process.}
  \label{fig:schematic}
\end{figure*}

The entire apparatus consists of three main sections: a cold-atom beam source, a science chamber with ultracold atoms, and a loadlock chamber (see Fig.~\ref{fig:schematic}(a)). The cold-atom beam source (AOSense, Inc.) is separated from the science chamber by a mini UHV gate valve ---all UHV valves are from VAT Group AG. The science chamber is held at $1\times10^{-11}$~Torr with a 40 l/s ion pump (Gamma Vacuum) and two NEGs
(Gamma Vacuum) at 200 l/s pumping speed each. Inside the science chamber is a sample holder (Ferrovac GmbH) mounted with a steel bar and groove grabbers (Kimball Physics Inc.). Samples loaded into the loadlock are transferred to the science chamber with a wobblestick manipulator (Ferrovac GmbH). 

For ease of reference, the loadlock chamber is further organized into five subsections: {\em loadlock}, {\em UHV pumps}, {\em bridge}, {\em turbo}, and {\em venting}. Before transferring devices into the science chamber, the samples are loaded into the {\em loadlock} at atmospheric pressures and evacuated to $1\times10^{-11}$ Torr within 24 hours. The central component of the {\em loadlock} section is a small six-sided cube (Fig.~\ref{fig:schematic}(b)) with each connection referenced as (1-6) below. The cube is connected to (1) the science chamber through an all-metal UHV gate valve and (2) a {\em UHV pumping} section with a 40~l/s ion pump (Starcell Agilent) and two NEGs (Gamma Vacuum) at 200 and 300~l/s pumping speeds. The 1.5-inch internal diameter of the bellows between the science chamber and loadlock sets the maximum cross-section of the devices that can be loaded into the science chamber. A mini UHV gate valve (V1) between the {\em loadlock} and its {\em UHV pumping} section isolates the sensitive pumps while venting and prevents the need for re-conditioning and re-activating the NEGs. The cube also connects to (3) the turbo-molecular pump (TMP, Agilent) through an all-metal right-angle valve (V2) and an all-metal variable leak valve (V3). This design improves pumping conductance to the TMP while providing controlled venting and evacuation for sensitive devices, such as an optical nanofiber~\cite{Kestler2023}. The last three ports of the cube are connected to (4) the wobblestick manipulator, (5) a viewport for monitoring the sample after loading, and (6) a `blank' flange, which can be swapped out for a feedthrough flange depending on the nanophotonic device in use. The volumes of the {\em loadlock} and {\em UHV pumping} sections are 0.61~L and 0.73~L, respectively, ensuring ample vacuum pumping with the ion pump and NEGs upon complete isolation from the TMP.

The {\em bridge} and {\em turbo} sections interface the {\em loadlock} to the TMP through 6-inch bellows and two mini-UHV valves (V4, V7). The bellows minimize stress on the loadlock conflat flanges since the TMP is firmly mounted to another structure. The {\em turbo} section lies between V4 and V7 and consists of a 4-way cross and a residual gas analyzer (RGA), which is used to monitor leaks from the {\em venting} section when V4 is closed. Lastly, the {\em venting} section is isolated by another mini UHV gate valve (V5) and an all-metal right-angle valve (V6), which increases isolation and reduces the leak rate through the mini UHV valve. The {\em venting} section has KF high vacuum flanges and a capacitance diaphragm pressure gauge (Inficon Group AG) to monitor the venting process to atmospheric pressures and to avoid over-pressuring the chamber. The final valve (V8) is connected to the venting gas line and is used to fill the chamber during the venting procedure.

Various photonic devices require feedthroughs and additional cabling in-vacuum~\cite{Vetsch2010, Goban2012, Kestler2023}. The bottom flange of the {\em loadlock} can be easily exchanged during the procedure; however, the additional slack needed to reach the science chamber must be appropriately handled. Our design incorporates a vertically mounted linear bellows actuator (Lesker Inc.) at the top of the {\em loadlock} with a ring at the tip of the actuator. During loading, extra cabling is fed through the ring, and the actuator is retracted upwards, pulling the slack above the photonic device. The actuator is lowered when the device is transferred into the chamber, allowing the slack to reach the science chamber. The reverse procedure ensures the gate valve is free to close completely upon removing the device from the science chamber.

\section{Procedure}

In addition to the critical design of the {\em loadlock} chamber, carefully implemented venting and loading procedures contribute to the rapid cycling speeds in this work. Opening directly to atmospheric pressures can leave residual water on the chamber surface and large quantities of undesired atmospheric gases, resulting in longer evacuation times. A standard solution is to vent the chamber with a constant flow of dry nitrogen gas, which the NEGs can quickly pump. Unfortunately, this can limit the lifetime of the NEGs and their ability to adsorb hydrogen outgassing from the surrounding steel. Instead, we vent the chamber with ultra-high purity (UHP) argon to extend the NEG lifetime.

For additional cleanliness during the loading process, we also enclose the {\em loadlock} cube in a custom-built acrylic glovebox. Before sealing the glovebox, we place the new device, all the necessary tools for the exchange, and multiple annealed copper gaskets inside. The tools are cleaned with methanol and wrapped in UHV foil if the person loading needs to use them before handling the sample. The glovebox interior is also wiped down with methanol and ensured to be dust-free. We then flood the glovebox with UHP argon through an inline filter and maintain a positive pressure slightly below that of the chamber venting. If any air intake into the chamber occurs, it will come from the glovebox argon instead of the surrounding air. 

The procedure used to reach the rapid cycle times in this work is detailed in the three sections below and labeled (A) preparation, (B) loading, and (C) baking and cooling. Valves are noted as V\# as shown in Fig.~\ref{fig:schematic}(b,c).

\subsection{Preparation (5 hours)}

A crucial part of this procedure is how the device to be inserted is cleaned, assembled, and prepared.

\begin{enumerate}
    \item Install the glovebox around the cube so the viewport, `blank' flange, and wobblestick are contained inside the glovebox.
    \item Clean all the necessary tools for installing the device and removing and reconnecting the conflat flanges. We include wrenches, pliers, scissors, tweezers, silver-plated screws with washers, and multiple annealed copper gaskets. Sonicating the tools is encouraged if possible, but it was not used in this work. Wrap the tool handles in UHV foil as much as possible and place the tools in the glovebox.
    \item Assemble and clean the device, and place it in the glovebox.
    \item Seal the glovebox and ensure $\approx6$~Pa of positive pressure when flooded with argon.
\end{enumerate}

\subsection{Loading (1-2 hours)\label{sec:load}}

The loading procedure begins from UHV with a previous device already retracted from the science chamber to the {\em loadlock}, and V0 closed to isolate the science chamber from the {\em loadlock}. Keeping the {\em loadlock} section at UHV conditions requires V1 to remain open. Thus, the starting configuration for all the valves is V1, V4, and V7 opened, while all others are closed.

\begin{enumerate}
    \item Close V7 and take rate-of-rise data of the {\em turbo} and {\em bridge} sections. After 5 minutes of not pumping with the TMP, we perform an analog scan.
    \item Close V1 to isolate the NEG and ion pump while venting.
    \item Close V4 and turn off TMP.
    \item Begin the flow of argon to V8. We avoid over-pressuring by adding a `balloon' with a small hole ($\approx1$~mm diameter) along the argon tubing upstream of V8. When any section is vented above atmospheric pressure, the excess gas will flow back out of the balloon, ensuring the chamber does not exceed atmospheric pressure.
    \item Open V8 until the {\em venting} section is at atmospheric pressure on the capacitance diaphragm pressure gauge. The `balloon' should never deflate during this venting to ensure minimal atmospheric gases enter the chamber. Once the {\em venting} section is at atmospheric pressure, close V8.
    \item Open V6.
    \item Repeat the last two steps, alternating between V8 and V5, then V8 and V4, then V8 and V3/V2. Lastly, open V8 for continuous flow into the {\em loadlock} section. At this point, V0, V1, and V7 are closed, and all other valves are open.
    \item Begin flooding the glovebox with argon. Allow a few minutes for the argon to replace any air previously in the glovebox.
    \item\label{step:removeblank} Remove the `blank' flange from the bottom of the {\em loadlock} cube.
    \item Remove the old device from the {\em loadlock} cube and replace the `blank' flange with a feedthrough flange if necessary.
    \item\label{step:removews} Remove the wobblestick flange from the back of the {\em loadlock} cube and slide the wobblestick far enough back, leaving ample room to insert the new device.
    \item\label{step:insertdevice} Insert the new device into the wobblestick. Place a new annealed copper gasket on the wobblestick flange and slowly reconnect the wobblestick to the {\em loadlock} cube.
    \item Hand-tighten the wobblestick flange on the {\em loadlock} cube. The argon flow to the chamber must be reduced as the {\em loadlock} section is sealed. Tighten the wobblestick flange completely and close V8 to shut off the argon flow to the chamber.
    \item\label{step:tmpon} Open V7 and turn on the roughing pump and TMP.
    \item Close V6 and V5 and turn on RGAs. Perform a quick helium leak check by filling the glovebox with helium and monitoring the RGAs.
\end{enumerate}
Modifying the procedure from steps \ref{step:removeblank}-\ref{step:insertdevice} depending on the device requirements is straightforward. We also advise only opening one flange at a time, but space constraints can be unavoidable. If multiple flanges are opened simultaneously, we increase the argon flow to the chamber to keep the positive pressure flowing out of the {\em loadlock} chamber.

\subsection{Bake and Cooling (23 hours)\label{sec:bake}}

After a successful exchange and helium leak check, we re-attach the baking components to the {\em loadlock} section and begin baking. Bake preparation involves placing thermocouples in various locations on the steel chamber and then wrapping a thin layer of UHV foil to distribute the heat evenly. This is followed by wrapping tape heaters and two or three additional layers to minimize thermal losses during the bake. Leaving as much of the chamber prepared as possible significantly reduces our bake preparation time. The {\em UHV pumps}, {\em bridge}, {\em turbo}, and {\em venting} sections remain prepared for a bake throughout the entire loading process. 

\begin{itemize}
    \item Remove the glovebox from the {\em loadlock} cube and prepare the {\em loadlock} section for the bake.
    \item Heat the chamber at a rate of $\approx$0.75~C/minute up to $\approx$90~C. Our chamber continues to rise over the subsequent hour. 
    \item Degas both RGAs.
    \item Allow temperatures to settle around 110~C and adjust the necessary temperatures to even out any undesired gradients.
    \item Begin cooling the bake when water reaches $2\times10^{-8}$ Torr and argon reaches $< 1\times10^{-9}$ Torr.
    \item Once the chamber temperatures are around 35~C, take a 5 minute rate-of-rise scan of the {\em loadlock}, {\em bridge}, and {\em turbo} sections by closing V7.
    \item At chamber temperatures $\approx$ 30~C, open V1, and close V2 and V3. The {\em loadlock} is now isolated and pumped on the ion pump and NEGs to reach $1\times10^{-11}$ Torr.
\end{itemize}

\section{Results}

\begin{figure}[t]
    \includegraphics{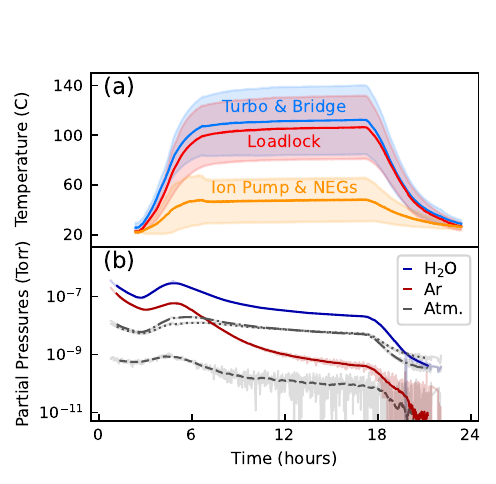}
    \caption{Pumping the loadlock chamber to $1\times10^{-11}$ Torr in 24 hours. (a)  The average temperature of the bake with the shaded region indicates the thermal gradient across each section. (b) Partial pressures of AMU 18 (water, blue solid line), AMU 40 (argon, red solid line), AMU 32 (oxygen, grey dashed line), AMU 28 (nitrogen, grey dash-dotted line), and AMU 44 (hydrocarbons \& carbon dioxide, grey dotted line) during the 24-hour pumping. Lighter colors indicate the raw data sampled every 3 seconds; darker colors denote a moving average over 5 minutes. The clean loading procedure results in similar pressures of water and argon, roughly an order of magnitude larger than the atmospheric gas pressures at the beginning due to the venting of the chamber with argon. A minimal pressure increase with a 110~C rise in temperature also indicates a clean procedure.}
  \label{fig:partialpressures}
\end{figure}

The procedure described in Section~\ref{sec:load} took a total of 1.2 hours to reach step~\ref{step:tmpon}, where we began timing the return from 1~atm to $1\times10^{-11}$~Torr shown in Fig.~\ref{fig:partialpressures}. Two critical indicators of the cleanliness of the loading protocol are the similar initial partial pressures of water and argon followed by a minimal increase in the partial pressures while heating the chamber. After the chamber reached a final temperature of around 110~C, the TMP efficiently pumped out the excess argon. We cooled the bake when the water reached a partial pressure of $2\times10^{-8}$~Torr at 110~C, achieving a two-order-of-magnitude drop upon reaching room temperature. 

While the chamber is heated, the small volume of the {\em loadlock} section increases thermal conductivity to the sample during the bake, efficiently heating the sample for more rapid cleaning. On a separate loading occasion, we successfully loaded and baked an optical ring resonator assembly, which included a thermistor for temperature monitoring. Baking the {\em loadlock} section at 70~C heated the in-vacuum sample to 50~C. Depending on the assembly materials, a modest bake of 110~C should not be problematic for many in-vacuum optics or glues. When we cannot reach the bake temperatures reported in this work, we have reached $1\times10^{-11}$~Torr within 36 hours.

\begin{figure}[t]
    \includegraphics{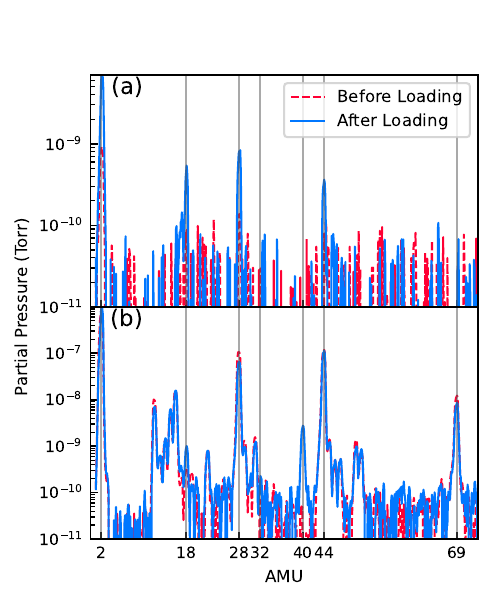}
    \caption{Scans from the Residual Gas Analyzer (RGA) before (red dashed line) and after (blue solid line) the loading procedure. `Before' data was taken with V4 and V7 open and all other valves closed at room temperature (22~C) with a total volume of 0.78~L. `After' data was taken with V2, V3, and V4 open and all other valves closed at 32~C and a total volume of 1.4~L. We scale the `after' scans by $V_{\rm{after}}/T_{\rm{after}}=1.23\times V_{\rm{before}}/T_{\rm{before}}$ relative to the `before' conditions.
    (a) An analog scan with the TMP operating indicates a clean chamber by reaching $<1\times10^{-9}$~Torr partial pressures on all gases but hydrogen. (b) An analog RGA scan after 5 minutes without pumping on the TMP (V7 closed) illustrates a similar composition of gases in the chamber before and after loading.}
  \label{fig:rgascan}
\end{figure}

The chamber temperature dropped below 40~C about 21 hours after turning on the TMP. We proceed to take two RGA scans with the Bridge RGA; one with the TMP pumping (V7 open, Fig.~\ref{fig:rgascan}(a)) and another scan five minutes after isolating the chamber from the TMP (V7 closed, Fig.~\ref{fig:rgascan}(b)). The isolated scan (V7 closed) indicates a similar composition of atmospheric gases before the loading procedure at AMU 28 (nitrogen) and AMU 32 (oxygen). These similarities also include AMU 44, 16, 15, 14, and 12, which result from the hydrocarbons at the high-temperature RGA probe. We also note that AMU 69 is present before and after loading as well as AMU 50 and 51, which we attribute to tetrafluorides in the {\em turbo} section since the scans lack the characteristic `unzipping' of hydrocarbons or mechanical pump oil. Furthermore, closing V4 instead of V7 shows no increase in AMU 69, 50, or 51. The main notable difference between the `before' and `after' conditions is the partial pressure of AMU 40 (argon), which is still low enough to be pumped by our ion pump upon isolation despite the two orders of magnitude increase after loading.

\begin{figure}[t]
    \includegraphics{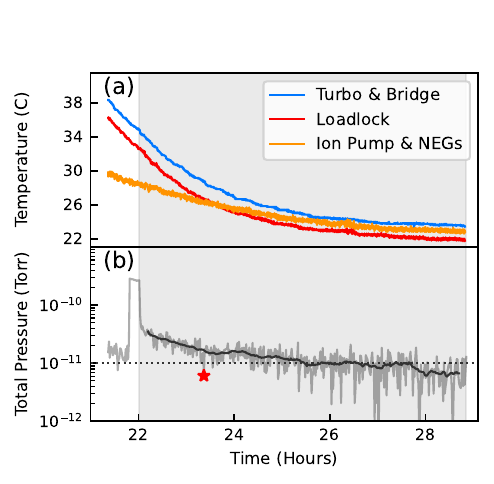}
    \caption{Total pressure reaching $<1\times10^{-11}$~Torr in the {\em loadlock} section. Both plots' light-grey shaded region indicates when the apparatus was fully isolated from the TMP. (a) Temperature decrease of the vacuum chamber cooling naturally from 38~C to room temperature. Steel retains heat, limiting our pump-down time. (b) Total pressure reaches $<1\times10^{-11}$~Torr within 24 hours. A red star notes when we could load the sample into the science chamber. The light-gray line indicates raw data from the ion pump controller sampled every minute. The dark solid line is a moving average of 10 minutes of data.}
  \label{fig:totalpressures}
\end{figure}

We isolated the {\em loadlock} section from the TMP when the chamber temperature reached $\approx$35~C, about 22 hours after turning on the TMP. Because we leave most of the vacuum chamber wrapped in preparation for a bake, the steel retains heat from the bake and takes additional time to cool from 35~C to room temperature. The continual drop in total pressure is consistent with the cooling of the {\em UHV pumps} section of the vacuum chamber (see Fig.~\ref{fig:totalpressures}), where we monitor the final pressure. We reach $<1\times10^{-11}$~Torr after 23.5 hours and continue to decrease pressure consistent with the decrease in the {\em UHV pumps} temperature.%

\section{Outlook}

Nearly all ultracold atom experiments require extensive optical setups and UHV pressures on the order of $1\times10^{-11}$~Torr for operation. Miniaturizing these experiments to a chip-scale device would significantly increase accessibility to quantum technologies. However, the path forward is iterative, and extensive prototyping will be necessary. Thus, there is an immediate need for rapid test systems. 

In this work, we present an ultra-high vacuum loadlock apparatus and procedure capable of loading chip-scale photonic devices at atmospheric pressures and returning to $<1\times10^{-11}$~Torr in less than 24 hours. The relatively small total volume directly results in rapid pumping speeds with a turbo-molecular pump and ultra-high vacuum pressures without chamber sputtering or cryogenics. The isolated ion pump and NEGs, as well as the choice to vent with argon, preserve the lifetime of the sensitive pumping equipment. 

The versatile design allows for loading various photonic devices, including those with vacuum feedthroughs, such as optical fibers and electrical wiring. We have successfully loaded optical nanofibers and optical ring resonators into the science chamber to be integrated with ultracold strontium atoms. 

\begin{acknowledgements}

We want to thank P. Lauria for valuable input into the design and initial construction of the apparatus. We also thank W. Brunner for help assembling the newest version of the system. We acknowledge the Office of Naval Research's support under Grant No. N00014-20-1-2693.

\end{acknowledgements}

\section*{Author Declarations}
\subsection*{Conflict of Interest}
The authors have no conflicts to disclose.

\subsection*{Author Contributions}
{\bf G. Kestler} contributed to the design, construction, data collection, and the written manuscript. {\bf K. Ton} contributed to the construction and data collection. {\bf J. T. Barreiro} contributed to the conceptualization, design, written manuscript, and funding procurement.

\section*{Data Availability Statement}

The data that support the findings of this study are available from the corresponding author upon reasonable request.

\section*{References}
\bibliography{24h_vacuum_bib}

\end{document}